\documentstyle[prl,aps]{revtex}
\let\section=\subsection  \let\subsection=\subsubsection

\def\be{\begin{equation}}
\def\ee{\end{equation}}
\def\bea{\begin{eqnarray}}
\def\eea{\end{eqnarray}}

\input{epsf}

\begin{document}

\begin{center}
{\large {\bf An Effective Field Model to High Temperature Superconductors:
Fitting of The Energy Gap to the Experimental Results of the $%
Bi_2Sr_2CaCu_2O_8$ }}\\[8mm]{E.C. Bastone \footnote{%
work partially developed at DCNat - Departamento de Ci\^{e}ncias Naturais,
FUNREI - Funda\c{c}\~{a}o de Ensino Superior de S\~{a}o Jo\~{a}o del Rei. Pra%
\c{c}a Dom Helvecio, CEP:36.300-000,S\~{a}o Jo\~{a}o del Rei - M.G. Brazil}
and P.R. Silva\\[5mm]{\small {\it Departamento de F\'{\i}sica, Instituto de
Ci\^{e}ncias Exatas,\\Universidade Federal de Minas Gerais,\\Belo Horizonte,
CEP 30.123-970, C.P. 702, MG, Brazil \\e-mail: erika@fisica.ufmg.br\\[8mm]}}}
\end{center}

\centerline{{\bf ABSTRACT}}
\begin{abstract}
\noindent
We use an effective field model (transverse Ising model) to describe the
dependence on the temperature of the energy gap of some two-dimensional $%
(2-D)$ superconducting systems. The order parameter of this model is put in
a direct correspondence with that of the Ising model. Then, we use the exact
relations for the spontaneous magnetization of the Ising model in some $2-D$
lattices as a means to fit the experimental results of $Bi_2Sr_2CaCu_2O_8 $
films obtained through reflectivity measurements by Brunel et al [Phys. Rev.
Lett. 66, 1346 (1991)] and the argon-annealed $Bi_2Sr_2CaCu_2O_8$ samples
investigated by Staufer et al [Phys. Rev. Lett. 68, 1069 (1992)] through
Raman scattering. The zero temperature energy gap $E_g(0)$ relations were
evaluated in various $2-D$ lattices and also in the $(3-D)$ simple cubic
lattice, for comparison. In the square lattice case we obtained $%
E_g(0)=3.52kT_c$, coincidentally the same number as the BCS result, and in
the triangular lattice case we get $E_g(0)=3.30kT_c$ in agreement with the
experimental findings of Brunel et al.\\[5mm]
\end{abstract}
Pacs number: 74.20.De, 74.72.Hs \\

As has been pointed out by Anderson $^{1}$ : the consensus is that there is
absolutely no consensus on the theory of high temperature superconductivity.
Anderson $^{2}$ attributes the novel phenomenology present on cuprates
materials to a second kind of metallic state, namely, the Luttinger liquid.
Schrieffer $^{1}$ has pursued the interplay between antiferromagnetism and
superconductivity, extending the BCS pairing theory beyond the Fermi-liquid
regime in terms of spin polarons or ''bags''. As pointed out by Cox and
Maple $^{3}$, the cuprates materials possess strong electronic correlations,
where by ''strong correlations'' we mean that the average interaction energy
substantially exceeds the average kinetic energy of the partially filled $3d$
state of the $Cu^{2+}$ ion. Therefore the discussion of what would be the
proper mechanism to describe the high temperature superconductivity in
cuprates becomes polarized between the BCS theory defenders as Schrieffer,
and the people who trust in the Hubbard model description as Anderson $^{1}$%
.This controversy suggest us that the proposition of new effective
hamiltonians, which does not enter in specific details of the mechanism,
could be an alternative way to look at this problem. Inspired in the
Gorter-Casimir two-fluid model of superconductivity $^{4}$, and also taking
into account the Blinc-de Gennes model $^{5,6}$ for the hydrogen-bonded
ferroelectrics, Gaona J. and Silva $^{7}$ have proposed the following
effective Hamiltonian as a means to describe some basic features of the
superconducting state: 
\begin{equation}
H=-\Omega \sum X_i-\frac 12\sum J_{ij}Z_iZ_j\text{.}
\end{equation}

The second term of (1) is a Ising-like term, where the operator $Z$ is
related to the wave function of the condensate of holes (electrons) and the
coupling $J_{ij}$ favors the pairing of holes at two different lattice
sites, contributing to the coherence of their wave functions below the
critical temperature $T_c$. In the first term of (1) the operator $X$ is
related to the wave function of the normal electrons (holes), where $\Omega $
represents the transverse field. This term, which gives dynamics to the
model, could represent a mimic for the motion of the free electrons (holes)
through the Fermi barrier, in close analogy with the tunneling of protons
through the double-well potential barrier in the KH$_2$PO$_4$-like
(hydrogen-bonded ferroelectrics) case (see $^5$).

We treat the dynamics of Hamiltonian (1), by applying the Random Phase
Approximation (RPA), where we replace a time dependent expectation value $%
<S_i>_t$ by a constant part $<S_i>$ - which is just the MFA expectation
value - plus a small time dependent deviation $\delta <S_i>e^{i\omega t}$
from the molecular field solution, to the Heisenberg equation of motion. The
solution of it in the first order approximation is $^{7}$:

\begin{equation}
\omega ^2=J_0^2<Z>^2+\Omega \left( \Omega -J_0<X>\right) \text{ .}
\end{equation}
This solution is constrained by the zero order approximation 
\begin{equation}
\left( \Omega -J_0<X>\right) <Z>=0\text{ .}
\end{equation}
Below $T_c$ , $<Z>\neq 0$ and by putting (3) into(2), leads to 
\begin{equation}
\omega _{\pm }=\pm J_0<Z>\text{ .}
\end{equation}
In (4) $\omega _{\pm }$ can represent a two level system, and we could treat
it as a gas of elementary excitations related to the hole (electron)
condensate. It seems natural to interpret the separation in energy of this
effective two-level system as the energy gap $E_g\left( T\right) $. We have
(in units where $\hbar =1$) 
\begin{equation}
2\Delta \left( T\right) =E_g\left( T\right) =\omega _{-}\left( T\right)
-\omega _{+}\left( T\right) =2J_0<Z>\text{.}
\end{equation}

Now, by using an idea introduced by Rummer and Ryvkin $^{8}$, since the
quasi-particles are non interacting fermions and their numbers are not
fixed, then the averaged occupancy of the levels $\omega _{\pm }\left(
T\right) $ is determined by the Fermi-Dirac distribution formula with zero
chemical potential. Then we have 
\begin{equation}
<n\left( \omega \right) >=\left[ e^{\beta \omega }+1\right] ^{-1}\text{,}
\end{equation}
where $\beta =1/kT$. By using the above distribution and associating a $%
\left| -\right\rangle $ eigenstate to the $\omega _{+}$ energy and a $\left|
+\right\rangle $ eigenstate to the $\omega _{-}$ energy, we obtain the
following relation for the superconductor order parameter $<Z>$, namely 
\begin{equation}
<Z>=\left( -1\right) \left\langle n_{+}\right\rangle +\left( +1\right)
\left\langle n_{-}\right\rangle \text{ ,}
\end{equation}
or alternatively we can write

\begin{equation}
<Z>=\tanh \left[ \left( \beta J_0<Z>\right) /2\right] \text{ .}
\end{equation}
Close to $T_c$, $<Z>$ goes to zero and we have from (8): 
\begin{equation}
2kT_c=J_0\text{.}
\end{equation}
At $T=0$, $<Z>=1$ and by putting (9) into (5) we obtain the energy gap 
\begin{equation}
E_g\left( 0\right) =2\Delta \left( 0\right) =2J_0=4kT_c\text{.}
\end{equation}

It is interesting to mention that Hao $^9$ have obtained this value for the
energy gap as one of the possible solutions of the BCS $^{10}$
self-consistency equation. Equation (8) can be put in a direct
correspondence with the self-consistency equation for the Ising model in the
mean field approximation (MFA), but with the temperature multiplied by a
factor of $2$ $^{11}$. Now let us consider the situation where the thermal
correlations close to $T_c$ are very important, so that the MFA is a bad
description to the order parameter behavior of the system. Indeed this seems
to be the case in some two-dimensional superconducting systems, as for
instance the $Bi_2Sr_2CaCu_2O_8$, where the order parameter were
experimentally determined through Raman scattering and reflectivity
measurements $^{12,13}$. The temperature dependence of the order parameter
of the Ising model in two-dimensions is exactly known $^{14}$. We can assume
that, for a two-dimensional ($2-D$) superconducting system where thermal
correlations can not be neglected, the order parameter can be mapped into
the exact relation for the spontaneous magnetization of the $2-D$ Ising
ferromagnetic model. However in this case and in the neighborhood of $T_c$,
the gas of elementary excitations approximation does not work quite well and
we must turn to the Boltzmann statistics. Now let us consider the exact
relation for the spontaneous magnetization of the Ising model in the square
lattice $^{14-16}$: 
\begin{equation}
m=\left[ \frac{2\tanh ^2\left( 2\beta J\right) -1}{\tanh ^4\left( 2\beta
J\right) }\right] ^{1/8}\text{.}
\end{equation}

In figures 1, 2 and 3, we compare eq. (11) with the experimentally
determined order parameter in $Bi_2Sr_2CaCu_2O_8$ single crystals,
investigated through Raman scattering by Staufer et al. $^{12}$ and in films
of the same material obtained through reflectivity measurements by Brunel et
al. $^{13}$. For comparison, we plot in the same figure the eq. (8) and the
BCS curve $^{10}$. We observe that the curve given by eq. (11) nicely fits
to the measurements of the energy gap performed at the $Bi_2Sr_2CaCu_2O_8$
films, a manifestly $2-D$ system. On the other hand, in the case of single
crystals, the adjust to curve (11) is good in the case of the essentially $%
2-D$ argon-annealed samples. In the oxygen annealed crystals the energy gap
can be described both by BCS theory as equivalently by eq. (8). In fitting
the experimental points to eq. (11), we have considered that: 
\begin{equation}
m=\frac{E_g\left( T\right) }{E_g\left( 0\right) }=\frac{2\Delta \left(
T\right) }{2\Delta \left( 0\right) }\text{,}
\end{equation}
and in the case of eq. (8) $m$ coincides with $<Z>$.

Now let us consider the exact result for the critical temperature of the
Ising model in the square lattice $^{14,15}$. We have 
\begin{equation}
\frac{kT_c}J=2.27
\end{equation}
which leads to the zero temperature energy gap 
\begin{equation}
2\Delta \left( T\right) =2J_0=2zJ=\frac{8kT_c}{2.27}=3.52kT_c\text{ ,}
\end{equation}
where we took $z=4$ (the coordination number of the square lattice). The
result we got in (14), is coincidentally the same as the BCS result and must
be compared with the experimental finding of $3.3kT_c$, obtained by Brunel
et al. $^{13}$. It must be emphasized that although the zero temperature
energy gap has been derived from a mean field analysis (eq. (5)), its use in
conjunction with the exact result for $T_c$ (eq.(13)) does not lead to
contradictions, once at $T=0$, we are far from the critical temperature $T_c$%
, where thermal correlations are important and the mean field treatment
substantially differs from the exact one.

Until now, we have compared the experimental measurements of the order
parameter of the $Bi_2Sr_2CaCu_2O_8$ with the exact result of the
spontaneous magnetization of the Ising ferromagnetic in the square lattice.
It would be interesting to pursue further on this subject, by comparing
within the same token, the zero temperature energy gap obtained by
considering other $2-D$ and $3-D$ lattices. Therefore in the following we
present the relations for the spontaneous magnetization (SM) of the Ising
model in some other $2-D$ lattices.

In the case of the honeycomb lattice $^{17,18}$ we have: 
\begin{equation}
m=\left[ \frac{1-2\sqrt{1-\tanh ^2\left( 2\beta J\right) }}{\tanh ^2\left(
2\beta J\right) \left[ 2-\tanh ^2\left( 2\beta J\right) -2\tanh ^2\left(
2\beta J\right) -2\sqrt{1-\tanh ^2\left( 2\beta J\right) }\right] }\right]
^{1/8}\text{.}
\end{equation}
For the triangular lattice case, the order parameter behavior is given by
the equation $^{15,19}$: 
\begin{equation}
m=\left[ \frac{2\tanh \left( 2\beta J\right) -1}{2\tanh ^3\left( 2\beta
J\right) -\tanh ^4\left( 2\beta J\right) }\right] ^{1/8}\text{.}
\end{equation}
In the case of the simple cubic lattice we do not have an exact relation for
the SM. However, fifteen years ago, one of the present authors $^{20}$ was
able to ''guess'' a relation for the SM of this $3-D$ lattice, which
reproduces very closely the series results of Fisher $^{21}$. This relation
is 
\begin{equation}
m=\left[ \frac{3\tanh ^2\left( 3\beta J\right) -1}{2\tanh ^3\left( 3\beta
J\right) }\right] ^{5/16}\text{.}
\end{equation}
In table 1 we present the energy gaps of various $2-D$ lattices. For sake of
comparison the $3-D$ simple cubic lattice is also quoted in this table.

Also is worth to notice that for the $2-D$ lattices, the order parameter
goes to zero at $T_c$ as $(T_c-T)^{1/8}$ $^{14-15, 17-19}$, while it behaves
as $(T_c-T)^{5/16}$ in the $3-D$ lattice $^{20}$ and as $(T_c-T)^{1/2}$ in
the BCS theory $^{10}$ (a mean field like behavior). In figure 4 we compare
the experimental data of Brunel et al $^{13}$ with the exact result for the
order parameter of the Ising model on the triangular lattice $^{19}$. We
again obtain a nice fitting of the experimental data to the theoretical
curve, with the bonus that the zero temperature gap of the theory reproduces
that experimentally observed. We would like also to comment that the
superconducting state of pairs of spinless quasiparticles was considered on
the basis of a model hamiltonian in MFA, by Safanov $^{22}$. One of the
results obtained by him is that the order parameter has a steeper rise with $%
(1-T/T_c)$ in the case of parastatistics than in the case of Fermi
statistics.

Finally, ultrafast dynamical optical response of $YBa_2Cu_3O_{7-\delta }$
was recently investigated by Stevens et al $^{23}$. They found that the
optical response is strongly peaked at $1.5eV$, and contains two distinct
components: one with a characteristic relaxation time of $\sim 5ps$, and a
long-lived component $(>10ns)$ which is consistent with localized
quasiparticle states at Fermi energy. For the slow component its
differential transmittance behaves with the temperature as $\Delta \tau
/\tau \propto \exp \left( -2\Delta _0/kT\right) $, where $2\Delta _0=3.5kT_c$%
. This corresponds to a thermal activated behavior with a fixed gap $2\Delta
_0$. We can interpret this experimental result of Stevens et al $^{23}$,
taking into account that in two-dimensional systems the gap only appreciably
deviates from its zero temperature value very close to $T_c$ (see figure 1),
and as we have determined in the square lattice case, this value is given by 
$2\Delta _0=3.52kT_c$.

The present research has been supported partly by CNPq-Brazil.

\[
\begin{tabular}{ccccccccc}
\hline
& Lattice &  &  & Transitions Temperature &  &  & Zero Temperature &  \\ 
&  &  &  & (values of $kT_c$ in $J$ ) &  &  & Energy Gap &  \\ \hline
& Honeycomb $(z=3)$ &  &  & $2.104$ $^{[17]}$ &  &  & $2.85kT_c$ &  \\ 
& Square $(z=4)$ &  &  & $2.27$ $^{[14]}$ &  &  & $3.52kT_c$ &  \\ 
& Triangular $(z=6)$ &  &  & $3.64$ $^{[19]}$ &  &  & $3.30kT_c$ &  \\ 
& Simple Cubic $(z=6)$ &  &  & $4.51$ $^{[21]}$ &  &  & $2.66kT_c$ &  \\ 
\hline
\end{tabular}
\]
\\[5mm]

Table 1: Zero temperature energy gap evaluated in this work for various $2-D$
lattices and for the simple cubic lattice $(3-D)$.\\[5mm]

\newpage 

\begin{figure}[h]
\vspace{0.8cm}
\centerline{\epsffile{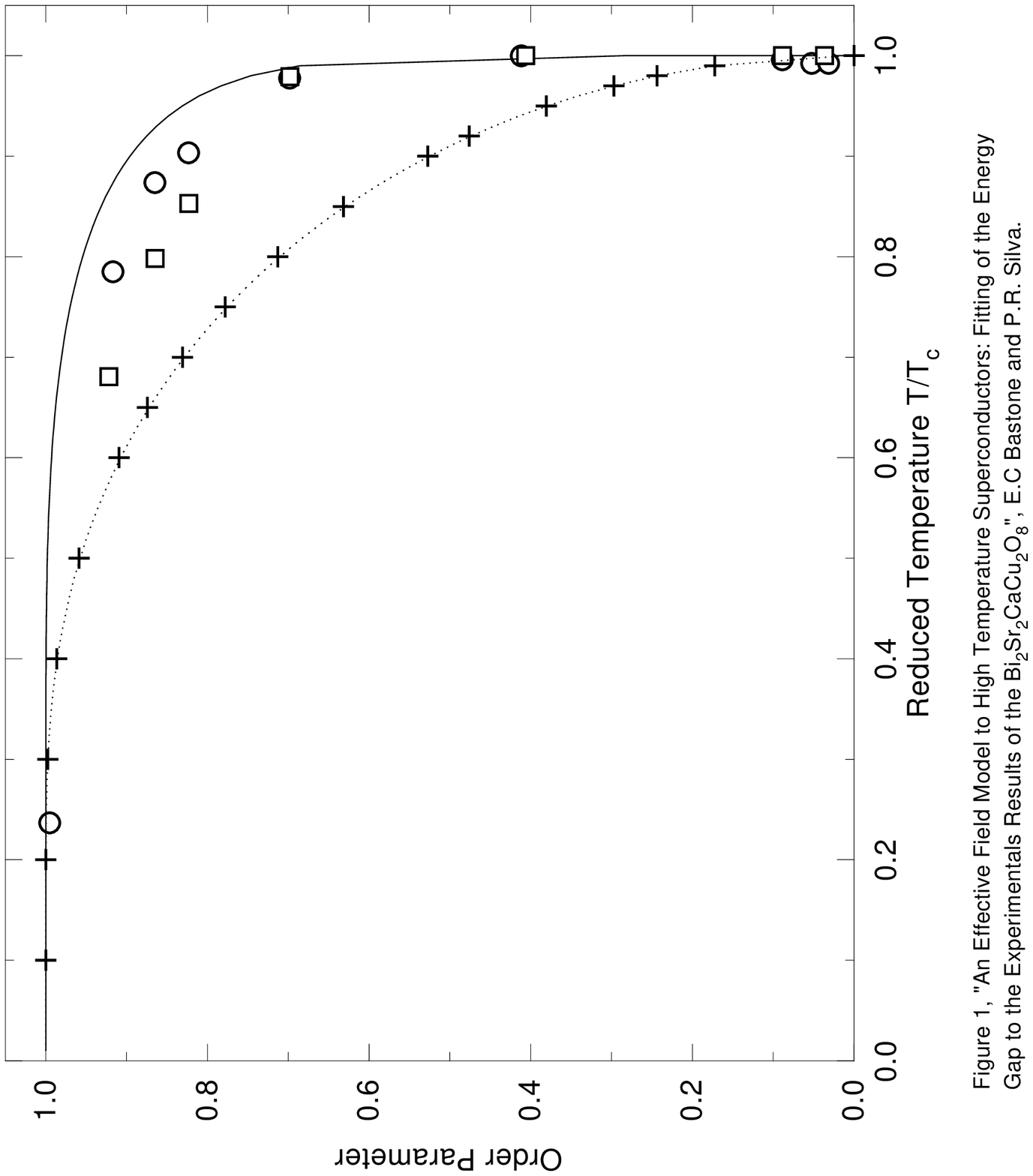} }
\vspace{0.8cm}
\label{double-well}
\end{figure}

\begin{figure}[h]
\vspace{0.8cm}
\centerline{\epsffile{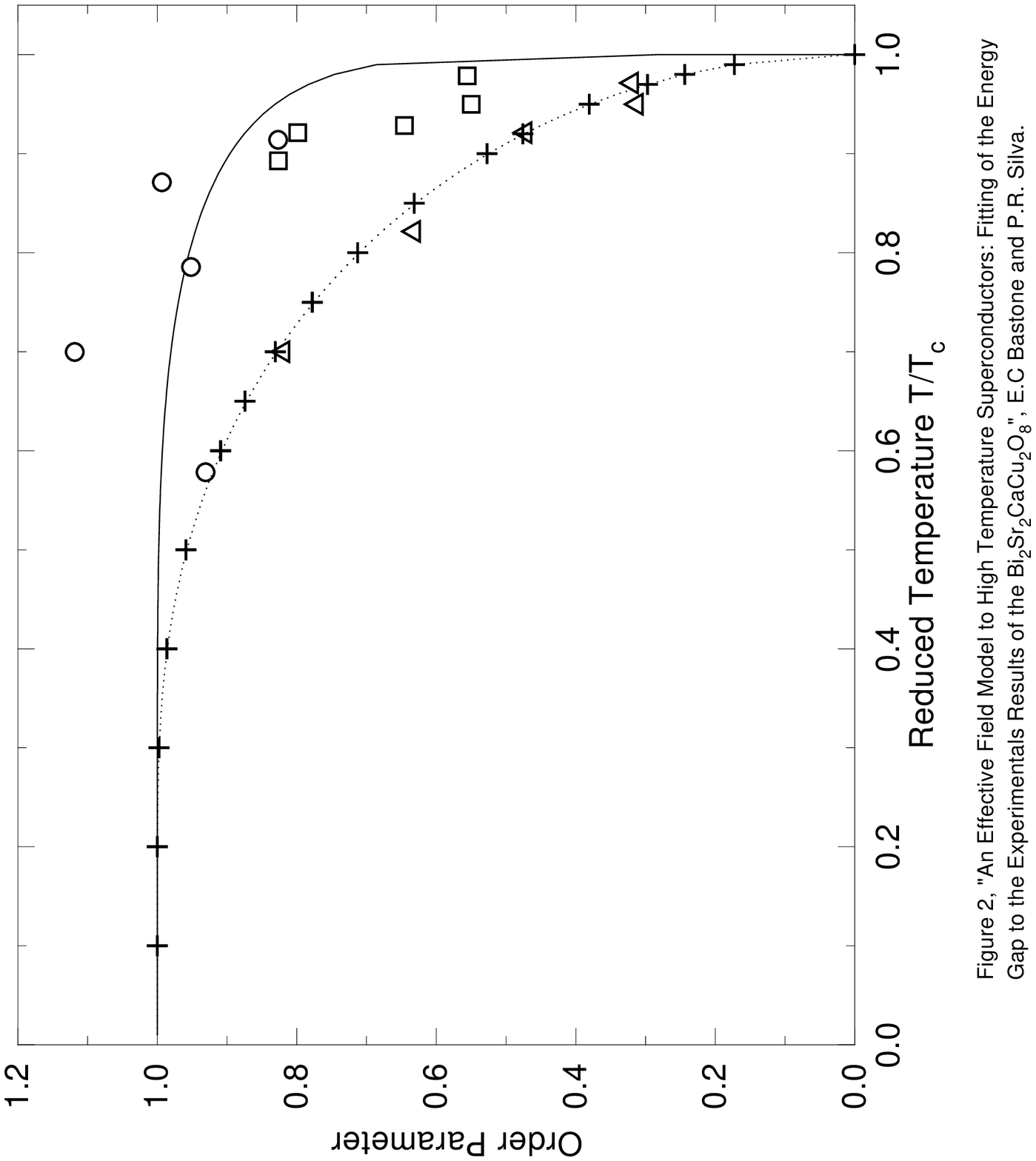} }
\vspace{0.8cm}
\label{splitting}
\end{figure}

\begin{figure}[h]
\vspace{0.8cm}
\centerline{\epsffile{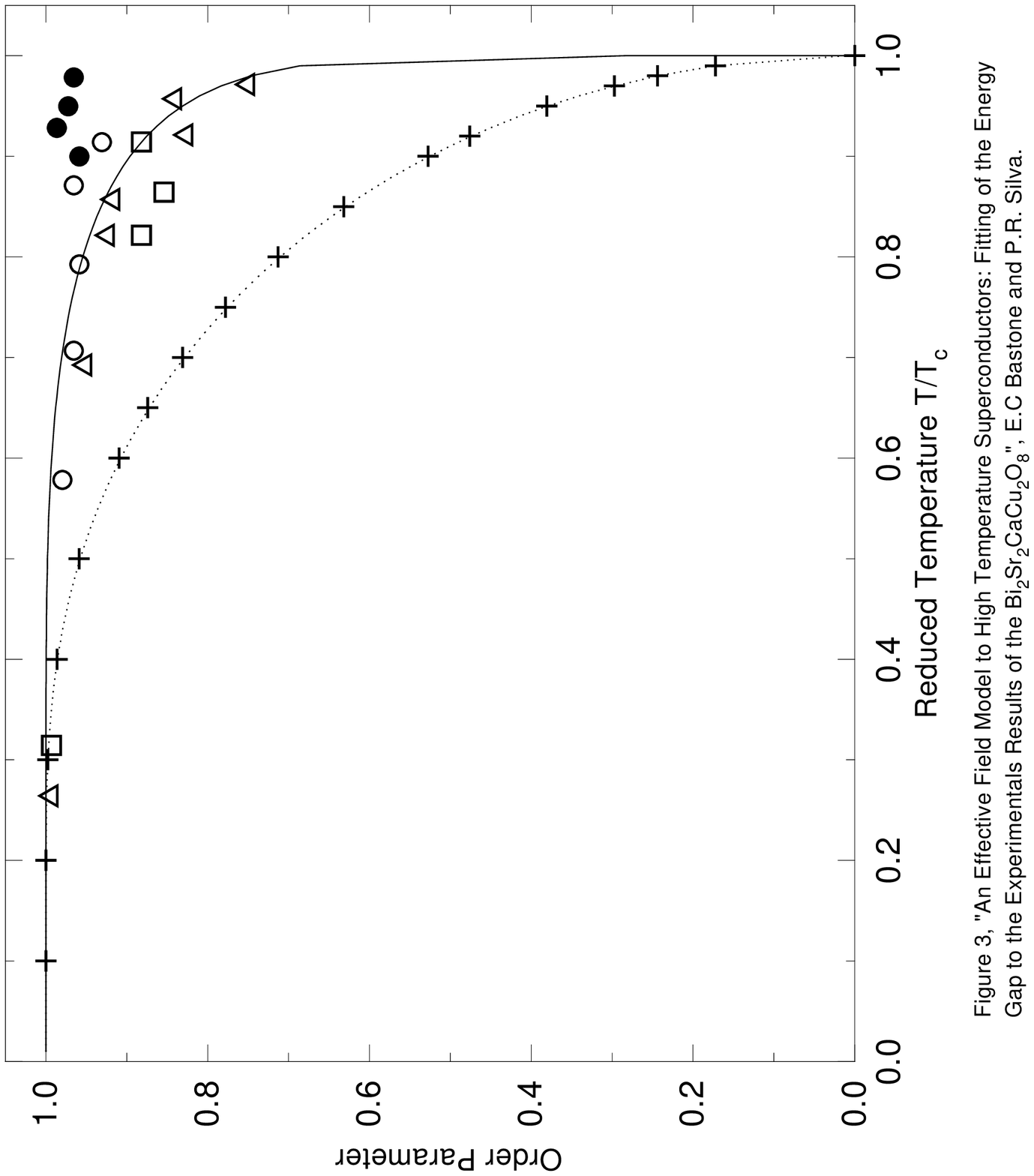} }
\vspace{0.8cm}
\label{resistivity}
\end{figure}

\begin{figure}[h]
\vspace{0.8cm}
\centerline{\epsffile{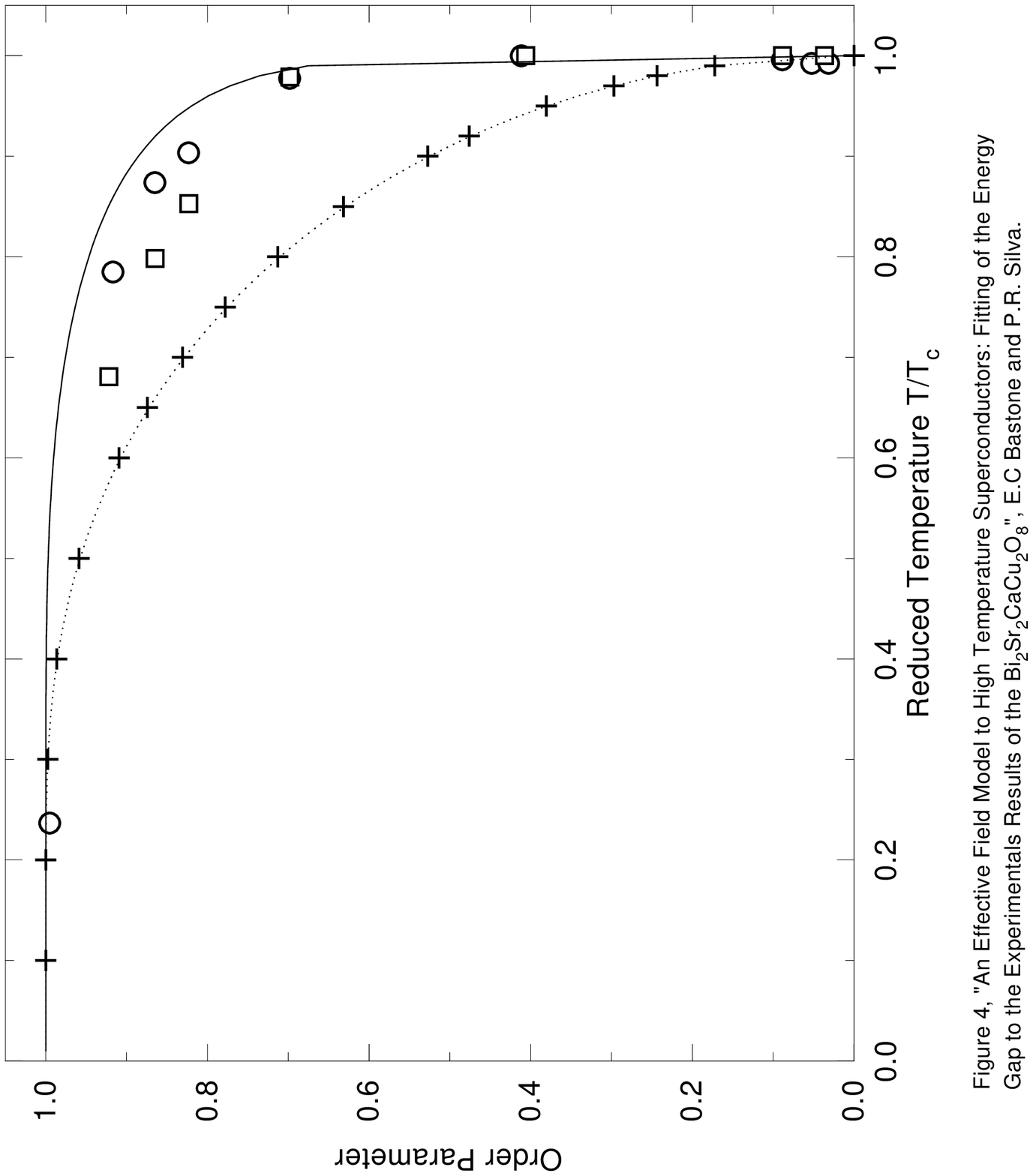} }
\vspace{0.8cm}
\label{mobility}
\end{figure}

\newpage

Figure Captions\\

Figure 1: Order parameter as a function of reduced temperature $T/T_c$. The
full line is the exact Ising curve for the square lattice. The points are
experimental results for superconducting films of $Bi_2Sr_2CaCu_2O_8$
obtained by Brunel et al $^{13}$: open circles are obtained at $T_c=87K$ and
open squares at $T_c=78K$. The dotted line represents the BCS theory and the
crosses are the MFA results (equation (8)).\\ 

Figure 2: Order parameter as a function of reduced temperature $T/T_c$. The
full line is the exact Ising curve for the square lattice. The points are
experimental results for superconducting samples of $Bi_2Sr_2CaCu_2O_8$
obtained by Staufer et al $^{12}$: open circles are as-grown annealed at $%
T_c=86K$, open squares are annealed in flowing $Ar$ at $T_c=86K$ and the
triangles are annealed in $O_2$ at $T_c=79K$. The samples are observed at $xy
$ polarization. The dotted line represents the BCS theory and the crosses
are the MFA results (equation (8)). \\ 

Figure 3: Order parameter as a function of reduced temperature $T/T_c$. The
full line is the exact Ising curve for the square lattice. The points are
experimental results for superconducting samples of $Bi_2Sr_2CaCu_2O_8$
obtained by Staufer et al $^{12}$: open circles are as-grown annealed at $%
T_c=86K$, solid circles are annealed in flowing $Ar$ at $T_c=86K$, the
triangles are annealed in $O_2$ at $T_c=79K$. The samples are observed at $xy
$ polarization. The squares are annealed in $O_2$ at $T_c=79K$ but observed
at $xx$ polarization. The dotted line represents the BCS theory and the
crosses are the MFA results (equation (8)). \\ 

Figure 4: Order parameter as a function of reduced temperature $T/T_c$. The
full line is the exact Ising curve for the triangular lattice. The points
are experimental results for superconducting films of $Bi_2Sr_2CaCu_2O_8$
obtained by Brunel et al $^{13}$: open circles are obtained at $T_c=87K$ and
open squares at $T_c=78K$. The dotted line represents the BCS theory and the
crosses are the MFA results (equation (8)).\\

\end{document}